\documentclass[%
 reprint,
 amsmath,amssymb,
 aps,
]{revtex4-1}

\usepackage{graphicx}
\usepackage{dcolumn}
\usepackage{bm}
\usepackage{color}

\begin{document}

\preprint{APS/123-QED}

\title{Dynamic and static analyses of glass-like properties of three-dimensional tissues}

\author{H. Nogucci}
\affiliation{%
Graduate School of Arts and Sciences, the University of Tokyo.
}%


\date{\today}

\begin{abstract}
The mechanical properties of cells, which influence the properties of the tissue they belong to, are controlled by various mechanisms.	
Bi et al. theoretically demonstrated that density-independent rigidity transition occurs in two-dimensional confluent tissues that consist of mechanically uniform cells.
They also analyzed the dynamical behavior of tissues near the critical point, which is geometrically controlled by `shape parameter'.
To investigate whether the behavior of three-dimensional tissues is similar to that of two-dimensional ones, we extend the model proposed by Bi et al. to a three-dimensional one both dynamically and statically.
The model reveals that the two mechanical states exist with a phase transition and has some similarities with those of glassy materials.
Scaling analysis is applied to the static model focused in the rearrangement viewpoint.
The results suggest that the upper critical dimension is also the same as the jamming transition.
\end{abstract}

\maketitle


\section{\label{sec:intro}introduction}

Cell rearrangement is frequently observed in confluent tissues and play particular roles during developmental processes, wound healing, and cancer metastasis \cite{Campbell2016}.
Because these processes have many common features,
they are categorized into Epithelial-Mesenchymal Transition (EMT).
Epithelial cells are tightly packed and located adjacent to each other, and mesencymal cells are less to each other.
EMT is a phenomenon where the mechanical properties of identical tissues changes from the epithelial to mesenchymal state at different time points, and the opposite phenomenon is called MET.
These macroscopic states originate from the mechanical properties of cells, such as cortical tension and intercellular adhesion, which are regulated by the state of gene expressions.

Two analogies can be applied to MET from the physical point of view.
One is phase transition from solid to liquid.
Epithelial tissues are mechanically rigid, while mesenchymal tissues are fluid-like.
However, the configuration of epithelial cells is irregular; their behavior is similar to that of glassy materials.
Most glassy materials undergo a jamming transition that is usually observed when the packing rate $\rho$ changes. Confluent tissues experience EMT, although the packing rate vis kept constant ($\rho = 1$) \cite{Berthier2011}.
The second analogy is a phase transition on collective motions of active matter \cite{Vicsek1995, Marchetti2013}.
Active matter consist of particles that move individually by consuming energy supplied from outside.
These examples have various length scales such as a flock of birds, groups of cells, and intracellular components.
They form dynamical collective orders by their cooperative motions.
Density phase transition has been intensively studied for active matters, while EMT occurs independently of density change.

EMT has been studied experimentally, particularly in two-dimensional systems.
To measure the distribution of forces on the edges between two contacting cells in developmental processes,
a Bayesian force inference was proposed \cite{Ishihara2012}.
Using this method, Sugimura et al. showed that mechanical anisotropy promotes cell packing with hexagonal ordering in Drosophila pupal wing \cite{Sugimura2013}.
The structural reconstructions of holes made by laser ablations in epithelial sheets were investigated in the context of wound healing \cite{Cochet-Escartin2014}.
Breast cancer cells show individual pulsating migrations in epithelial tissues owing to the mismatch of mechanical properties, which provides an insight into tumor progression \cite{Lee2012, Palmieri2015a}.
The analysis of collective cell motions in a tissue was performed by tracking individual cells and their glassy behavior, such as caging and dynamic heterogeneity, was reported \cite{Schotz2013}.

Recently, Bi et al has showed the existence of a new type of phase transition, observed in two-dimensional tissues consisting of mechanically uniform cells.
They studied a model with a phenomenological energy functional originating from cellular shapes. 
The energy functional $E$ is the total energy of an individual cell $E_i$ which is described as
\begin{equation}
E_i = K_{A} (A_i - A_{0})^2 + \xi P_i^2 + \gamma P_i,
\end{equation}
where $A_{i}$ and $P_{i}$ are the mean area and perimeter of the cell indexed $i$.
$K_{A}$, $\xi$, and $\gamma$ represent the cell's elastic constant in two-dimensional systems, active contractility driven by the cytoskeletons present in cells, and interfacial tension between contacting cells, respectively.
$A_{0}$ denotes the optimal cell area in the isolated situation.
They introduced a `shape parameter' $\tilde{P} = -\gamma/2\xi$, which is an optimal cell perimeter in the energy ground state, and found that rigidity transition occurs around $\tilde{P}_0 \sim 3.81$.
At this value, the optimal cell shape is a regular pentagon, and the energy cost of cell rearrangement vanishes when $\tilde{P}$ is larger than $\tilde{P}_0$.
They also showed that collective cell motions drastically change around the optimal value of the parameter.
While individual cells move diffusively if $\tilde{P}$ and the magnitude of self-propelling velocity $v_{0}$ are large, some are caged by their surrounding cells and the collective motion is heterogeneous elsewhere. 

Despite their work, the behaviors of three-dimensional tissues have not been investigated at present.
The existence of rigidity transition in three-dimensional cases remains to be elucidated.
In this study, we extend the energy functional $E$ to the three-dimensional system.
The individual cell energy $E_i$ is described as
\begin{equation}
E_i = K_{V} (V_i - V_{0})^2 + K_{A} (A_i - A_{0})^2,
\end{equation}
where $V_i$ and $V_0$ means a cell volume and its optimal value in a single cell system.
$K_V$ represents the cell's elastic constant in three-dimensional systems \cite{Schwarz2013}.

In this studied, the collective cell behaviors in three-dimensional confluent tissues are investigated by using dynamical and static models.
Model settings of dynamical cell motions are explained in Sec.~\ref{sec:model1}, and their results are described in Sec.~\ref{sec:result1}, where the phase transition of collective motions is similar to glass transition, depending on the shape parameter introduced later.
To investigate the static behavior around the transition, the cell rearrangement energy is measured and analyzed with a scaling method described in Sec.~\ref{sec:model2}.
Lastly, the conclusions and discussion are given in Sec.~\ref{sec:conclusion}.

\section{\label{sec:model1}model settings of dynamical cell behaviors}
\subsection{Model Equations}
To describe collective cellular motion considering cell shapes, the Voronoi cell model is adopted used by Bi et al. in Ref.~\cite{Bia}.
In this model, the position and direction of a cell $i$ are represented as $\textbf{x}_i ( = (x_i, y_i, z_i) )$ and $\textbf{p}_i ( = ({p}_{{x}_i}, {p}_{{y}_i}, {p}_{{z}_i}))$, respectively.
The position is denoted by the cell center, and the direction is denoted by a unit vector pointing from the tail to head of the cell.
The shape of the cell is approximated through graph representation made by 3D Voronoi tessellation of $\left\{\textbf{x}_i \right\}$.
This approximation implies that the shape must represent a convex polyhedron and that there should be no vacant space in the system.
Throughout this paper, the boundary condition is set to be periodic.
Euler's polyhedron formula leads to the following relation: $\#f-\#e+\#v= 2$, where $\#f$, $\#e$, and $\#v$ denote the total number of faces,  edges, and vertices of a single cell.
A vertex connects three edges if degeneracy is ignored, which results in the relation $2\#e = 3\#v$.
These relations reveal that $\#e$ and $\#v$ is directly derived, if $\#f$ is known. 
 
We assume that the interacting forces acting on each cell $i$ is described as ${\textbf{F}}_{i} = - \nabla_{i} E$, which has the same form as Ref.\cite{Bia}.
Cellular motion also consists of self-propulsion whose magnitude is assumed to be constant $v_0$.
With the overdamped equation of motion, the cell position $\textbf{x}_i$ is governed by two terms:
\begin{equation}
{\dot{\textbf{x}}}_{i} = \mu {\textbf{F}}_{i} + v_0 \textbf{p}_i
\end{equation} 
The directions of the cells are assumed to perturbate randomly within its paralell plane:
\begin{equation}
{\dot{\textbf{p}}}_{i} = \nu {\bm{\eta}}_{i} (t) \times {\textbf{p}}_{i},
\end{equation}
where $\mu$ is the mass of each cell divided by the drag coefficient, and $\nu$ denotes the moment of inertia divided by the rotational drag coefficient.
The random vector ${\bm{{\eta}}}_{i} (t) ( = ( {{\eta}_i}_x (t), {{\eta}_i}_y (t), {{\eta}_i}_z (t) ) )$ obeys the following statistics: 
\begin{equation}
\langle {{\eta}_{i}}_{k} (t) {{\eta}_j}_{k'} (t)\rangle = 2D {\delta}_{ij} {\delta}_{kk'} {\delta}(t-t'),
\end{equation}
where ${\delta}_{ij}$ and ${\delta}_{kk'}$ are the Kronecker delta on the cell indices and the component indices of the vector, respectively, and ${\delta}(t-t')$ is the Dirac delta function of the time variables.
$D$ is the magnitude of the directional change of cell motions.

\subsection{Rescale and Parameter Settings}
Energy functional can be expressed as follows:
\begin{equation}
E_i = K_{V} {V_0}^2 ({\tilde{V}}_i - 1)^2 + K_{A} {V_0}^{4/3}({\tilde{A}}_i - {\tilde{A}}_{0})^2, \label{eq:3des}
\end{equation}
where ${\tilde{V}}_i = V_i / V_0$ and ${\tilde{A}}_i = A_i / {V_0}^{2/3}$ are the rescaled volume and surface area, respectively.
While ${\tilde{A}}_{0}$ originates from the mechanical properties of the cells,
it determines the optimal surface area per unit volume that minimizes the shape energy functional if the shape of all cells are confined to be equal, convex, and isotropic [Table ~\ref{tab:valueref}].
In this paper, this parameter is referred to as `shape parameter'.

To simplify the mathematical form, we select the unit length of the system as ${V_0}^{1/3}$ and set ${V_0} = 1$.
The system size $L$ is set to be $L = 6 {V_0}^{1/3}$, and the total number of cells $N$ is set to be $N = 6^3$ so that the packing ratio $NV_0/{L^3}$ is equal to unity.
The energy ratio is defined as $r = K_A / (K_V {V_0}^{2/3})$ and we fix $r = 1$ in Sec.~\ref{sec:result1}.
Some parameters regarding the cell's properties are also fixed: $\mu = 1$, $\nu = 1$, and $D = 0.1$. 
Cell position is randomly partitioned for initial configuration.

Numerical simulations are performed until $t = 22000$, whose time unit is $1/(\mu K_V V_0)$ with a fixed step size $\Delta t = 0.1$.
The statistical values for all parameter regions, which are referred to in Sec.~\ref{sec:result1} and \ref{sec:model2}, are calculated by averaging $5$ different samples that starts from different initial conditions.

From the geometrical viewpoint, the shape parameter ${\tilde{A}}_{0}$ is related to the Kelvin problem:
How can space be partitioned into cells of equal volume with the least surface area?
In case the shape of all cells is identical, a truncated octahedron is believed to be the optimal shape \cite{Weaire1994}.
\begin{table}[h]
\caption{\label{tab:valueref}
Regular shape and the corresponding value of the shape parameter ${\tilde{A}}_0$. 
}
\begin{ruledtabular}
\begin{tabular}{lr}
\textrm{shape}&
\textrm{${\tilde{A}}_0$}\\
\colrule
sphere & 4.836\\
icosahedron & 5.148\\
dodecahedron& 5.312\\
truncated octahedron & 5.315\\
octahedron & 5.719\\
cube & 6\\
\end{tabular}
\end{ruledtabular}
\end{table}

\section{\label{sec:result1}dynamical cell behaviors}

\subsection{Diffisive and Sub-diffusive Collective Motions}
Two distinct collective motions are observed on changing the parameter values $v_0$ and ${\tilde{A}}_0$.
In case the values of both $v_0$ and ${\tilde{A}}_0$ are small, cell rearrangement is hardly observed and collective motion is as slow as glass.
A fast and fluid-like collective motion is observed when the values of $v_0$ and ${\tilde{A}}_0$ are large.

\begin{figure}[h]
\includegraphics[width=8.6cm]{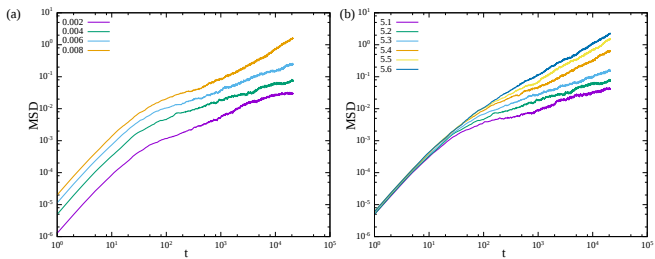}
\caption{\label{fig:AvMSD} MSD of cell trajectories.
(a) MSD for different $V_0$ values with fixed $\tilde{A}_0 ( = 5.1)$.
(b) MSD for different $\tilde{A}_0$ values with fixed $V_0 ( = 0.004)$.
}
\end{figure}

To characterize these motions quantitatively, the mean squared displacement (MSD) of the cell trajectories is measured, as shown in Fig.~\ref{fig:AvMSD}.
$MSD(t)$ is defined as $MSD(t -  t') = \sum_{i = 1}^{N} |\textbf{x}_i (t) - \textbf{x}_i (t')|^2/N, t' = 2000$, where $|\cdot|$ is $L$--$2$ norm with the periodic boundary.
For all parameter regions, $MSD(t)$ is proportional to $t^2$ when $t$ is small.
This indicates that ballistic motion is dominant in this time scale.
In case of a large $t$ value, $MSD(t)$ is proportional to $t$, if both $v_0$ and $\tilde{A}_0$ are large. However, $MSD(t)$ is proportional to $t^d,~(d<1)$, when either diffusive or sub-diffusive motions are observed for a long period depending on the parameter values.

To investigate as to why $MSD(t)$ shows different behaviors for the two time scales, 
we first divide the two time scales by the length scale of MSD(t).
For cell rearrangement, cells must move as far as $\sim 0.01{V_0}^{1/3}$, which is comparable to the edge length of a single cell.
Self-propulsion is dominant before cells move farther away, as shape force or diffusion is dominant after cell rearrangement.

\begin{figure}[h]
\includegraphics[width=5cm]{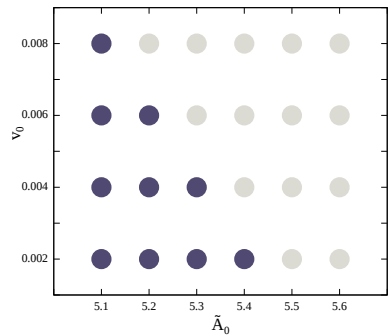}
\caption{\label{fig:PD} Phase diagram for the two parameters $v_0$ and $\tilde{A}_0$.
Sub-diffusive collective motions are observed in the parameter region filled with dark-blue points, while
diffusive motions are observed in the gray-colored region.
}
\end{figure}

Second, we characterize long-term collective motions.
The unit of self-diffusivity $D_0$ is introduced as $D_0 = v_0^2/(3D)$, and the magnitude of self-diffusivity $D_s$ is measured as $D_s = \lim_{t \to \infty} MSD(t)/(6t)$.
We practically measured $D_s$ by averaging $MSD(t)/(6t)$ for a value of $t$ that satisfies $MSD(t) > 0.01$.
If $M_s/M_0$ is larger than the threshold that originate from noise floor, the collective motion is regarded as diffusive; otherwise, it is considered to be sub-diffusive.
The threshold value is set to be 0.05.
The phase diagram shown in Fig.~\ref{fig:PD} is obtained using this criterion.

Diffusive collective motions in the long term are observed when the magnitude of self-propulsion $v_0$ is high and the shape parameter $\tilde{A}_0$ is large, while sub-diffusive motions are observed if the value of both these parameter is small.
The critical point $\tilde{A}_{0*}$ in the limit $v_0 \to 0$ is larger than 5.4, although the critical point for the regular packing of a truncated octahedra shows $5.31$.
This point is of particular interest because, above this point, cell dynamics is purely dominated by the force originating from the shape energy functional, except for the noise effect, and cells can freely rearrange.
The exact critical point and behaviors near this point are both discussed in a later section.

\subsection{Similarities with Glassy Materials}

Next, we consider the properties of sub-diffusive motions $(d < 1)$.
This problem can be replaced by `How are these motions similar to that of glassy materials?'

The first reason could be `caging', a phenomenon where particles cannot move as they are surrounded by their nearest neighbors for a long period of time \cite{Berthier2011}.
\begin{figure}[b]
\includegraphics[width=5cm]{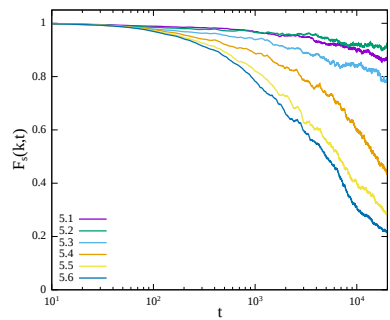}
\caption{\label{fig:sis}Self-intermediate scattering function ($F_s (k,t), ~ k = \pi/r_0, t_0 = 2000$) for different $\tilde{A}_0$ with fixed $v_0$ value $(v_0 = 0.004)$.}
\end{figure}
This phenomenon is represented by the self-intermediate scattering function $F_s (k,t)$.
It is defined as $F_s (k,t) = \langle  e^{i {\bf k}\cdot \Delta{\bf r}(t) }\rangle$, where $\Delta{\bf r}(t)$ denotes the difference in the position of the cell at the start time $t_0$ and the measured time $(t_0 = 2000)$ and $\langle \cdot \rangle$ represents the average over all the cells.
Figure. ~\ref{fig:sis} shows the value of $F_s (k,t)$ after averaging the angles of ${\bf k}$ for different parameters. 
The magnitude of ${\bf k}$ is fixed so that $F_s (k, 0) \equiv 1, ~(k = \pi/r_0)$, where $r_0$ is the averaged nearest position of contacting cells for each cell. 
If caging occurs, the value of $F_s(k,t)$ is kept near unity for a longer period of time.
For a fixed value of $v_0 = 0.004$, $\tilde{A_0} < 5.3$.
Elsewhere, the function approaches zero with the progress of time, indicating that the structure of the tissues is completely relaxed within the endtime of numerical simulations.

\begin{figure}[h]
\centering
\includegraphics[width=8.6cm]{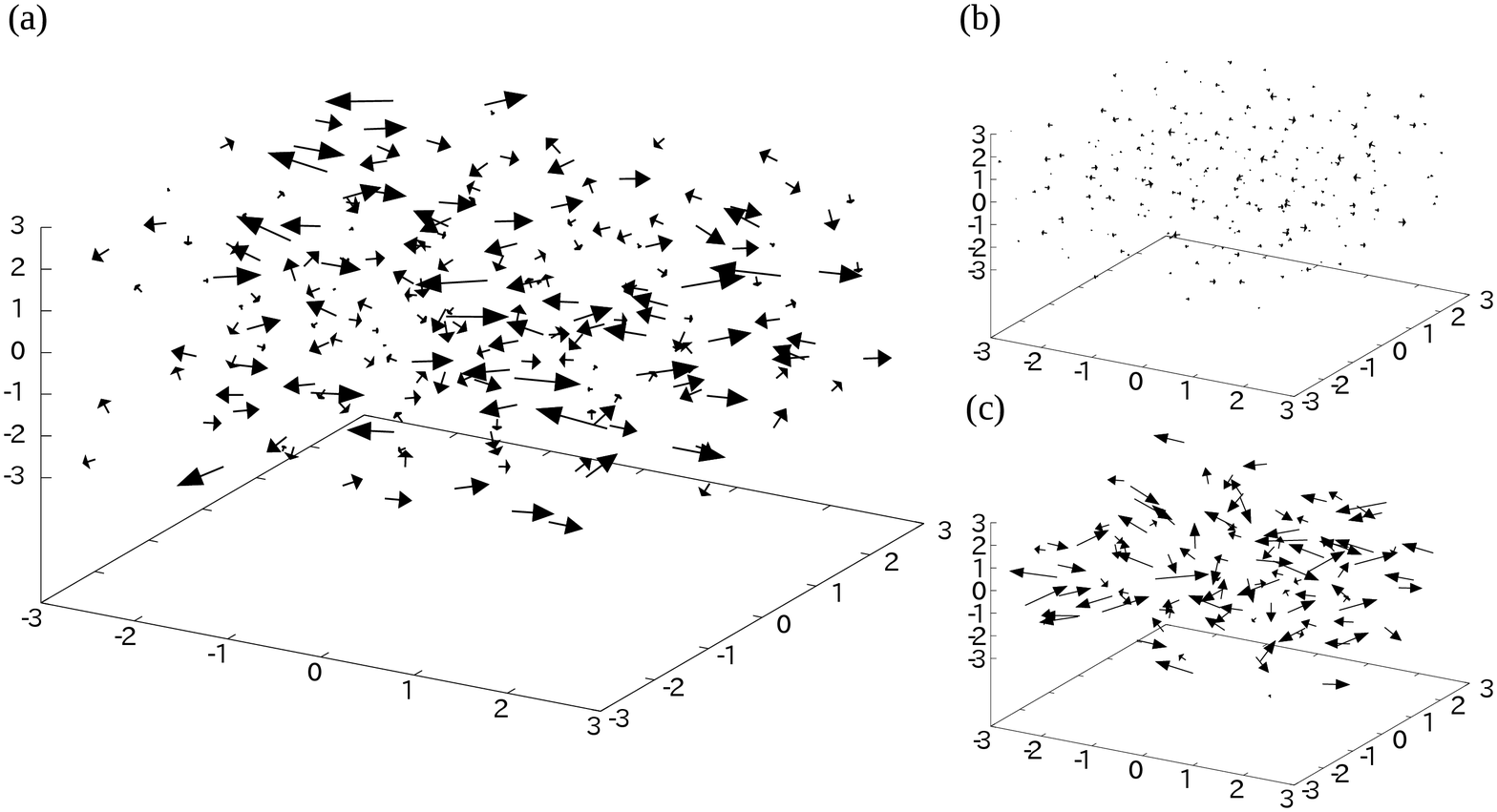}
\caption{\label{fig:hetero}Migration vectors of cells for three different parameter sets.
Duration is set from $t = 21000$ to $t = 22000$.
Snapshots are shown
in (a) with a diffusive parameter set near the transition boundary line($\tilde{A}_0 = 5.4, ~ v_0 = 0.004$),
in (b) with a sub-diffusive parameter set far from the boundary line ($\tilde{A}_0 = 5.1, ~ v_0 = 0.002$), and
in (c) with a diffusive parameter set far from the boundary line ($\tilde{A}_0 = 5.6, ~ v_0 = 0.008$), respectively.
}
\end{figure}

The second reason is dynamic heterogeneity\cite{Berthier2011}.
Migration vectors of the cells are shown in Fig.~\ref{fig:hetero}.
It is defined as a vector pointing from the starting position to the finish one and duration is taken with $10^3$ time scale $(t = 21000$ -- $22000$).
In the parameter region in that collective motions are diffusive but that is near the transition boundary from the sub-diffusive region, some cells move for the long distance but the others stay in small domains for long term.
This indicates that dynamical heterogeneity is also detected near the diffusive--sub-diffusive transition boundary line, suggesting that the dynamics of three-dimensional tissues are similar to those of glass.

\subsection{Analysis of the Individual Cell Shape}

\begin{figure}[h]
\includegraphics[width=5.5cm]{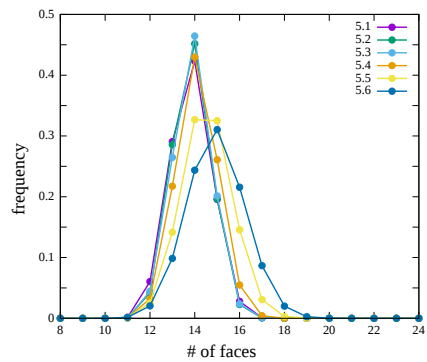}
\caption{\label{fig:faces}Distributions of the polyhedra for different parameters $\tilde{A}_0$ with $v_0 = 0.004$ averaged in the duration $t = 21000$--$22000$.}
\end{figure}

To understand the relationship between the individual cell shape and collective motions,
the distributions of cell shapes are measured.
In the adopted model system, cell shape is approximated as a convex polyhedron.
Figure.~\ref{fig:faces} shows how the number of faces for each cell is distributed for different $\tilde{A}_0$ with fixed $v_0 ( = 0.004)$.
At the parameter region where the collective motion is sub-diffusive, the average number of the faces is $14$.
As $\tilde{A}_0$ is larger, the system shows diffusive collective motion,
where the average number of faces is larger than $15$ and its variance is also larger than that with $\tilde{A}_0 < 5.3$.

Common phenomena are observed in the two-dimensional tissues in that the increasing shape index triggers the `sub-diffusive'-to-`diffusive' transition of the tissue \cite{Bia}.

\begin{figure}[h]
\includegraphics[width=8.6cm]{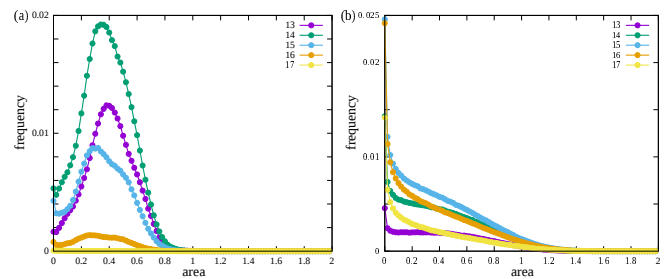}
\caption{\label{fig:areadis} Joint distributions of faces that belong to the $n$-faced polyhedoron with a different area value
with (a) sub-diffusive parameter set ($\tilde{A}_0 = 5.1$, $v_0 = 0.004$) and
with (b) diffusive parameter set ($\tilde{A}_0 = 5.6$, $v_0 = 0.008$).
}
\end{figure}

The lattices of regular truncated octahedrons are the global solution for the Kelvin problem discussed in Sec.~\ref{sec:model1}, and they also have 14-sided faces, which supports the hypothesis that the average shape of the individual cell observed in the sub-diffusive collective motions is the regular truncated octahedron.
In two-dimensional tissues, however, the critical point corresponds to the value of the shape parameter that represents the regular pentagon, although the value for the regular hexagon is the global solution for the energy minimum states of the tissues.
 We then compare the actual cell shapes to investigate whether the typical shape of the cells in the three-dimensional tissues is regular truncated-octahedron.
A truncated-polyhedron has $6$ regular squares and $8$ regular hexagons as its faces; the area when its volume is $1$ is $0.1984$ for the square faces and $0.5155$ for the hexagon faces, respectively.
Figure.~\ref{fig:areadis} shows the joint distribution for faces that belong to the $n$-faced polyhedron and have specific area value.
The distribution for the 14-faced polyhedron showing sub-diffusive motion shows a single peak with a value $\sim 0.4$ as the area value [Fig.~\ref{fig:areadis}(a)], indicating that the shapes of the cells are not similar to regular truncated tetrahedrons.
The lattice of the regular truncated polyhedron is the global solution for the Kelvin problem; however, the polyhedron does not have an isotropic shape while both model equations and the periodic boundary cube do not have the mechanism to break the symmetry of isotropy for the single cell.

On the other hand, the area $0$ peak is always found for all faces showing diffusive motion [Fig.~\ref{fig:areadis}(b)], indicating that cell rearrangement occurs for any cell shape.
Diffusive collective motions originate from free cell rearrangements.

\section{\label{sec:model2}critical point and scaling behaviors of cell rearrangement energies}

Phase transition from sub-diffusive collective motions to diffusive collective motions in the large time scale was investigated in the previous section; however, the value of the critical point $\tilde{A}_{0*}$ in the limit $v_0 \to 0$ remains unknown.
The next goal is to determine the value and study the critical behaviors near it by ignoring self-propulsion.
To determine the value of $\tilde{A}_{0*}$, the energies of cell rearrangements should be measured.
Hereafter, the measurement method is explained focusing on the parameter dependency only for the shape parameter $\tilde{A}_0$ and energy ratio $r$.

\subsection{Measurement Method}

To investigate the energetic properties of cell rearrangement, we introduce a static model.
In addition to the dynamical model in Sec.~\ref{sec:result1}, Eq.~\ref{eq:3des} is adopted as the energy functional originating from the cell shape constraint.
First, one of the states with the energy local minima is achieved from the random initial configuration of the cell positions with the gradient descent method.

\begin{figure}[h]
\includegraphics[width=8.6cm]{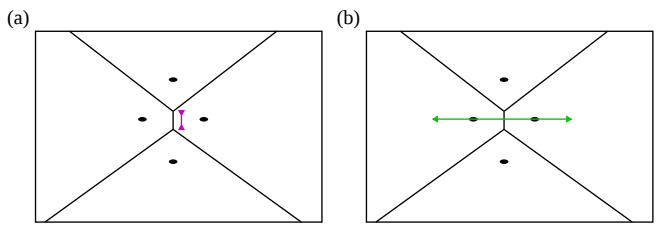}
\caption{\label{fig:fsidea}The schematic view of (a) the measurement method of the CVM model in Ref.~\cite{Bi2014} and (b) that of the SPV model used in this paper.
Only a two-dimensional scenario is expressed for simplicity.}
\end{figure}
Using the final configuration that reaches the energy local minima, we measure the rearrangement energy for contacting pairs of the cells. 
Figure.~\ref{fig:fsidea} schematically expresses the measurement difference from Ref.~\cite{Bi2014}.
For simplicity, we consider the cases of two-dimensional systems.
In Ref.~\cite{Bi2014}, a cellular vertex model is adopted and the variables are the positions of cell vertexes; therefore, rearrangement is generated by shortening the edge length of the contacting cells to zero.
In this paper, on the other hand, the SPV model is adopted and variables are the positions of cell centers; therefore, the length of the edge cannot be controlled.
Instead to operate the length, cell centers are moved in the opposite direction to each other because the pair will become unconnected after several iterations of this operation.
In the case of the three-dimensional system, contacting edges are replaced into contacting faces, while both the process and its efficiency remain unchanged.
The detailed procedures are explained in Appendix.~\ref{sec:AppA}.

\subsection{Critical Point and Scaling Behavior}
\begin{figure}[h]
\includegraphics[width=5.5cm]{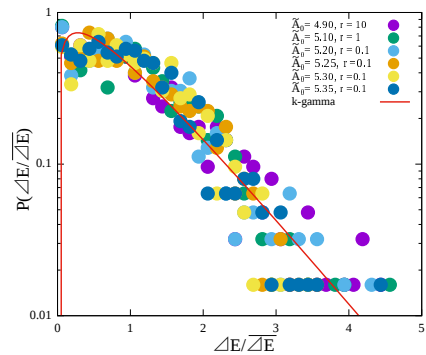}
\centering
\caption{\label{fig:ediffdis}The distributions of the rearrangement energy rescaled by the division of its average for different parameter combinations.
$200$ faces are selected for each simulation.
The red line shows the $k$-gamma distribution with $k = 1.38$.}
\end{figure}

The energy of a cellular rearrangement is marked $\Delta E$.
Figure.~\ref{fig:ediffdis} shows the distributions of $\Delta E$ rescaled by its average over the sample faces $\overline{\Delta E}$ for many parameter sets.
These can be well fitted with $k$-gamma distribution $\bigl( p(x) = k^k x^{k-1} \exp{-kx}/(k-1)! ,~(k = 1.38~(\pm 0.01) \bigr)$.
This distribution emerges due to the maximization of entropy with the constant packing ratio and it is
 observed in many kinds of disordered systems \cite{Aste2008, Newhall2011}.

\begin{figure}[bh]
\includegraphics[width=8.6cm]{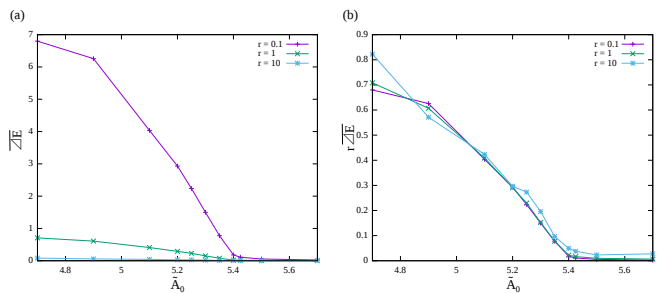}
\caption{\label{fig:ediffs}$\overline{\Delta E}$ for various parameter sets $\tilde{A}_0$ and $r$.
(a) The averaged energy of cell rearrangement $\overline{\Delta E}$.
(b) The rescaled averaged energy $r\overline{ \Delta E}$.}
\end{figure}

\begin{figure}[h]
\includegraphics[width=8.6cm]{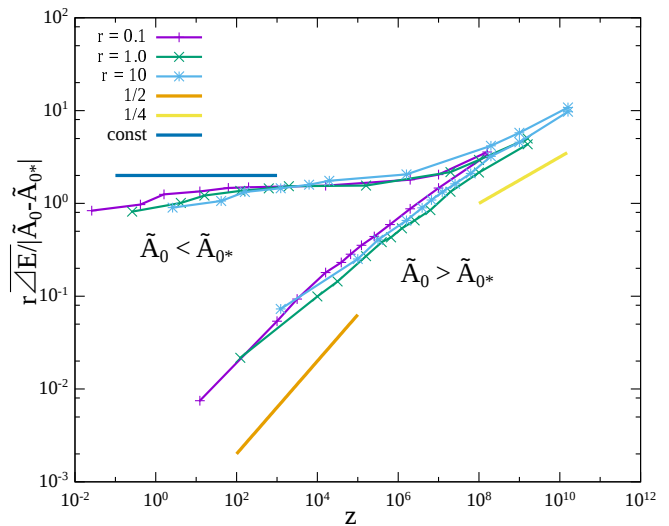}
\caption{\label{fig:scaling}
Scaling function with the values $(\tilde{A}_{0*}, \beta, \Delta) = (5.410, 1, 4)$.
Three slope lines are drawn to compare the universal curves with the fitted values of the critical exponents.
}
\end{figure}

$\overline{\Delta E}$ for different parameter sets is shown in Fig.~\ref{fig:ediffs}(a).
Depending on the magnitude of the energy rate $r$, the finite values of $\overline{\Delta E}$ are different among the region where $\tilde{A}_0$ is small.
After rescaling them by multiplying $r$, three curves agree with a single form [Fig.~\ref{fig:ediffs}(a)].
As measured here, the critical point $\tilde{A}_{0*}$ exists within the range from $5.4$ to $5.5$.
The rescaled rearrangement energy can be the candidate of the order parameter to classify two phases.
To examine the hypothesis, a scaling is adapted to the data, following the method performed by the previous study \cite{Bi2014}.
In the Ising model, magnetization $m$ is expressed with magnetic field $h$ and temperature difference from the critical temperature $T-T_c$.
In analogy with the relation in $(m, h, T - T_c)$ for spin statistical physics, $(r\overline{ \Delta E}, r, \tilde{A}_0 - \tilde{A}_{0*})$ should obey the scaling relation
\begin{equation}
r \overline{ \Delta E} = |\tilde{A}_0 - \tilde{A}_{0*}|^{\beta} f_{\pm}\left( \frac{r}{|\tilde{A}_0- \tilde{A}_{0*}|^{\Delta}} \right), \label{eq:scaling}
\end{equation}
where $\Delta$ is the crossover scaling critical exponent, and
$f_{+}$ and $f_{-}$ are the two branches of the crossover function whose sign added at its subscript corresponds to that of $(\tilde{A}_0- \tilde{A}_{0*})$.
$z$ is defined as $z = r/{|\tilde{A}_0-\tilde{A}_{0*}|}^{\Delta}$, which represents the crossover scaling variable.
The exponent $\beta$ represents a following relation in the limit $z \to 0$:  $r \overline{ \Delta E} \propto |\tilde{A}_0 - \tilde{A}_{0*}|^{\beta}$.
At the critical point, the two branches merge as $f_{+} = f_{-} = z^{\beta/\Delta}$
By changing various values of $\tilde{A}_{0*}$ and fitting $(\beta, \Delta)$ with them, the best fit for the scaling relation (\ref{eq:scaling}) is obtained by taking $(\tilde{A}_{0*}, \beta, \Delta) \sim (5.410, 1, 4)$ [Appendix.~\ref{sec:AppB}].
With this set of values, the scaling function is obtained for $z$ and is collapsed to a universal curve [Fig.~\ref{fig:scaling}].
As seen in Sec.~\ref{sec:result1}, the value $\tilde{A}_{0*}$ is not equal to that for the regular truncated octahedron.
In the branch where cell rearrangement is highly suppressed by finite energy barriers, its height is scaled as $\overline{ \Delta E} \propto K_V {V_0}^2 {(\tilde{A}_0 - \tilde{A}_{0*})}$, while it is described as $\overline{ \Delta E} = r^{{\beta/\Delta}-1}$ in the limit $z \to \infty$.

The values of $\beta$ and $\Delta$ are the same as those in the two-dimensional systems shown in Ref.~\cite{Bi2014}.
This correspondence beyond the difference of the dimension is also present in many types of systems that exhibit Jamming transition.

\section{\label{sec:conclusion}conclusion and discussion}

In this paper, the mechanical properties of the three-dimensional tissues were investigated as an extension for the research carried out by Bi et al.
It was shown that collective motions in the long term have two types, diffusive and sub-diffusive, which emerge depending on both the self-propulsion $v_0$ and on the shape parameter $\tilde{A}_0$.
Sub-diffusive collective motions contain caging cells and dynamic heterogeneity, which are observed in the parameter region near the transition boundary line.
These are common features as reported in glassy materials.
In analyzing the individual cell shape statistically, it is found that the shape at the phase transition point does not correspond to a regular truncated octahedron and instead, it is more isotropic.
 To measure the critical value $\tilde{A}_{0*}$ in the SPV model, rearrangement energy $\Delta E$ whose average obeys the scaling relation is used (\ref{eq:scaling}).
 The critical exponents are the same as those of the two-dimensional tissues, indicating that the upper critical dimension of the tissue is $2$, while the jamming transition of the glassy materials is believed to have the same upper critical dimension.
 
 Now, we briefly discuss possible relevance of our results to cell biology.
 First, EMT occurs in three-dimensional tissues, which will help understand various types of biological phenomena.
 Next, some tissues such as skin and trachea can be regarded as psuedo-two-dimensional because there are a few stacks of cells along an axis.
 If the upper critical dimension of the tissue is $2$, the tissues are similarly considered to exhibit glassy properties because both two- and three-dimensional tissues also show these properties \cite{Bi2014, Bia}.

We do not consider the dynamics of the mechanical properties of the single cell, such as elastic coefficients $K_V$ and $K_A$ and other kinds of interactions coupling $v_0$ and $\mathrm{p}_i$.
Containing their dynamics could lead to another type of collective motion, which may have a relation with chemotaxis and planar cell polarity.
We assume that tissues only consist of cells with uniform mechanical properties; however, breaking this assumption may also trigger interesting phenomena.

Since the shape index $\tilde{A}_0$ can be measured, the results of the paper were tested through experiments to assess the mechanical properties of the three-dimensional tissues.
Furthermore, measurements of the shape index for all cells with a three-dimensional image may unveil some anomalous events such as cancer metastasis.

\subsection{Note Added:}
 After the completion of the manuscript, we noted the preprint  \cite{Merkel2017}, where a similar problem is treated.
 Although we determined that the critical point does not correspond to the global solution of the energy functionals of the tissues, we showed the scaling properties of the energies of cell rearrangement and phase transition, which are yet to be reported.
\begin{acknowledgments}
I would like to thank Prof. K. Kaneko for the beneficial discussions and advice on some topics that relate to this paper.
\end{acknowledgments}

\appendix
\section{\label{sec:AppA}gradient method to measure rearrangement energies}
The gradient method is executed in two steps to measure the energies.
The first step is to set up the system that is dropped into a state with the energy local minima, and the second step is to measure the rearrangement energy after picking up a contacting pair of the cells.
 
To justify whether the whole system reaches energy local minima, the gradient method is executed for all numerical simulations and repeated until either the value of the total energy difference between current configuration and the previous one $|E(s-1)-E(s)|$ divided by the current total energy $|E(s)|$ becomes smaller than $10^{-5}$ or $2\cdot10^4$ MD steps $(s)$ are counted.

The method is also used during the forced rearrangement of the cells.
After contacting cells are forced to stay away at a constant distance determined by one-tenth of the initial distance between them, the method is applied until $\cdot10^3$ MD steps are counted.

\section{\label{sec:AppB}fitting the critical exponents}
As the computational cost is large, performing finite scaling analysis is difficult.
Instead, we fit $\tilde{A}_{0*}$ by changing the value as $\tilde{A}_{0*} = 5.40, 5.405, \dots , 5.45$.
Since the best fit curve is obtained by inserting $\tilde{A}_{0*} = 5.410$, we set up three different assumptions $(\tilde{A}_{0*} = 5.405, 5.410, 5.415)$ and fit other exponents $(\beta, \Delta)$.
The relation $\overline{\Delta E} \sim (\tilde{A}_{0*}-\tilde{A}_{0})^{\beta}$ in the limit $z \to 0$ for the branch of $\tilde{A}_{0} <\tilde{A}_{0*}$ is used to fit the value of $\beta$ and the relation $\overline{\Delta E} \sim r^{\beta/\Delta-1}$ in the limit $z \to \infty$ is used to fit the value of $(\beta/\Delta-1)$.

Table.~\ref{tb:fitdata} shows the fitted values using different values of $\tilde{A}_{0*}$.
Although the value of $\tilde{A}_{0*}$ is changed, the values of the fitted exponents do not change drastically around $\beta \sim 1.1$ and $\beta/\Delta-1 \sim -0.76$. This is how we conclude $(\tilde{A}_{0*}, \beta, \Delta) \sim (5.410, 1, 4)$.

\begin{table}
\begin{ruledtabular}
\begin{tabular}{c|c|c}
assumed value & fitted value & fitted value  \\
of $\tilde{A}_{0*}$ & of $\beta$ & of $\beta/\Delta-1$ \\ \colrule
5.405 & $1.108 (\pm 0.217)$ & $-0.751(\pm0.049)$ \\
5.410 & $1.109 (\pm 0.219)$ & $-0.769(\pm0.053)$ \\
5.415 & $1.105 (\pm 0.221)$ & $-0.795(\pm0.053)$ \\ 
\end{tabular}
\end{ruledtabular}
\caption{\label{tb:fitdata}Fitting $\beta$ and $(\beta/\Delta-1)$ by assuming the various values of $\tilde{A}_{0*}$.}
\end{table}



\end{document}